\documentstyle[aps,epsfig,subfigure,graphicx]{revtex}
\newcommand{\ket}[1]{\left| {#1} \right\rangle}
\newcommand{\nuc}[2]{\mbox{${}^{#1}${#2}}}
\newcommand{\parenth}[1]{\left({#1}\right)}
\newcommand{\brackets}[1]{\left[{#1}\right]}
\newcommand{\degree}{^{\circ}}
\newcommand{\eq}[1]{equation (\ref{#1})}
\newcommand{\eqn}[1]{(\ref{#1})}

\newlength{\figwidth}
\begin{document}

\author{H K Cummins\dag and J A Jones\dag\ddag}
\title{Use of Composite Rotations to Correct Systematic Errors\\ in NMR
Quantum Computation}
\address{\dag Centre for Quantum Computation,
Clarendon Laboratory, Parks Road, Oxford OX1 3PU, UK}
\address{\ddag Oxford Centre for Molecular Sciences, New Chemistry
Laboratory, South Parks Road, Oxford OX1 3QT} \maketitle

\begin{abstract}
We implement an ensemble quantum counting algorithm on three NMR
spectrometers with \nuc{1}{H} resonance frequencies of 500, 600
and 750 MHz. At higher frequencies, the results deviate markedly
from naive theoretical predictions. These systematic errors can be
attributed almost entirely to off-resonance effects, which can be
substantially corrected for using fully-compensating composite
rotation pulse sequences originally developed by Tycko. We also
derive an analytic expression for generating such sequences with
arbitrary rotation angles.
\end{abstract}

\section{Introduction}
Quantum computers are information processing devices which operate
by---and exploit---the laws of quantum mechanics, giving them the
potential to tackle problems which are intractable using classical
computers \cite{shor:1999}. Although there has been a great deal
of interest in the theory of quantum computation, actually
building a general purpose quantum computer has proved extremely
difficult. The multiparticle coherent states on which quantum
computers rely are extremely vulnerable to the effects of errors,
and so considerable effort has been expended on tackling random
errors introduced by decoherence, including methods of error
correction \cite{shor:1995,steane:1996} and fault tolerant
computation \cite{steane:1999}.  Comparatively little effort,
however, has been directed at the issue of systematic errors,
arising from reproducible imperfections in the apparatus used to
implement quantum computations.

It makes sense to address systematic errors, however, because many
of them can be eliminated relatively easily. It is convenient to
visualise one spin quantum states as three-dimensional vectors,
where the $x$, $y$, and $z$ components are given by decomposition
of the state into a linear combination of Pauli spin matrices. In
this picture, unitary operations are interpreted as rotations of
the Bloch vector on a sphere. The sensitivity of the final state
to rotational imperfections can be much reduced by replacing
single rotations with composed rotations. Ordinarily replacing one
erroneous operation by several erroneous operations will increase
the overall error, but this isn't necessarily the case as
rotations are non-linear, and the non-linearity opens the
possibility of errors cancelling one another.

It is worth stressing the distinction between improved
experimental technique, quantum error correcting codes
\cite{steane:1999}, and the use of composite pulses. Improved
experimental technique minimises conditions which lead to data
errors. Error correcting codes diagnose errors (however obliquely)
and use this knowledge of the errors to explicitly reverse them.
Composite pulses prevent errors in the first place by reducing the
impact on the system of conditions which cause systematic data
errors, without actually eliminating these conditions. Composite
pulses cannot, however, correct random errors, as they rely on the
reproducible nature of systematic errors to correct them.

\section{Systematic errors in NMR quantum computers}

In the last few years nuclear magnetic resonance (NMR) techniques
\cite{ernst:1997} have been used to implement small quantum
computers \cite{cory:1996}. Several simple quantum algorithms have
been implemented using NMR, such as Deutsch's algorithm
\cite{deutsch:1986,jones:1998c,chuang:1998b} and Grover's
algorithm \cite{grover:1997,chuang:1998,jones:1998a}. The
comparative ease with which such experiments have been performed
reflects, in part, the fact that more conventional NMR
experiments, used throughout the molecular sciences, themselves
rely on the generation and manipulation of multi-spin coherent
states, and so techniques for handling them are highly developed.

Despite this pre-existing experimental sophistication, NMR quantum
computers are nonetheless subject to systematic errors.
Implementing complex quantum algorithms require a large network of
logic gates, which within an NMR implementation entails even
longer cascades of pulses. In this case small systematic errors in
the pulses (which can be largely ignored in conventional NMR
experiments) accumulate and become significant.

Interactions between individual spins and between spins and the
environment are mediated by RF pulses, which are applications of
an RF field with phase $\phi$ (in the rotating frame)
\cite{ernst:1997} for some duration $\tau$. In the ideal case, the
pulse will drive the Bloch vector through an angle $\theta$ about
an axis orthogonal to the $z$-axis and at an angle $\phi$ to the
$x$-axis.  The rotation angle, $\theta$, depends on the rotation
rate induced by the RF field, usually written $\omega_1$, and the
duration of the pulse, $\tau$. In practice the RF field is not
ideal, and this leads to two important classes of systematic
errors, usually referred to as pulse length errors and
off-resonance effects.

Pulse length errors arise either when the length of the pulse is
set incorrectly, or (more commonly) when the RF field strength
deviates from its assumed value, so that the rotation angle
achieved deviates from its theoretical value. This effect is most
commonly observed within NMR as a result of inhomogeneity in the
applied RF field; in this case it is impossible for all the spins
within the sample to experience the same rotation angle.

Off resonance effects arise from the use of a single RF source to
excite transitions in two or more spins which have different
resonance frequencies.  This is done for a variety of practical
reasons.  For example, in conventional NMR experiments, which seek
to analyse molecular systems, there will be a large number of
different transition frequencies, whose exact values are unknown
at the start of the experiment.  Thus the only practical approach
is to use a single RF source, with sufficient power that it can
excite transitions across a wide range of frequencies.  In quantum
computation experiments the transition frequencies are known
beforehand, but for a variety of reasons the use of a single RF
source for each type of atomic nucleus remains the most practical
approach (different sources \emph{are} used for different nuclei,
such as \nuc{1}{H} and \nuc{13}{C}).

Composite pulses are widely used in NMR to minimize the
sensitivity of the system to pulse-length and off-resonance
errors, by replacing direct rotations with composite rotations
which are less sensitive to such effects. However, conventional
composite pulse sequences are rarely appropriate for quantum
computation because they usually incorporate assumptions about the
initial state of the spins or introduce phase errors.  These
conventional sequences were initially derived by considering the
trajectories of Bloch vectors during a pulse, and this approach is
only useful if the starting point of the Bloch vector is known.

QC requires fully-compensating (type A) composite pulse sequences
\cite{levitt:1986}. Fortuitously, a number of such sequences were
developed fifteen years ago by Tycko \cite{tycko:1983}.  While
these pulse sequences do not offer quite the same degree of
compensation as is found with some more conventional sequences,
the compensation which does occur is effective whatever the
initial position of the Bloch vector.  As such sequences are
rarely (if ever) required within conventional NMR experiments,
these sequences have previously found no experimental use.  They
are, however, ideally suited to quantum computations.

\section{Example: NMR quantum counting}

Systematic errors are an issue with many types of quantum
algorithms, but we will focus on a counting algorithm which can
involve particularly long pulse trains. Counting the number of
items matching a search criterion is a well-known computer science
problem, and an efficient quantum counting algorithm based on a
variation of Grover's quantum search has been implemented using
NMR techniques\cite{jones:counting:1999}.

Consider a match function $f(x)$ which maps $n$-bit binary strings
to a single output bit. In general there will be $N=2^n$ possible
input values, of which $k$ will yield a match, $f(x)$=1. The
object of the quantum counting algorithm is to estimate $k$ by
estimating an eigenvalue of the Grover iterate $G=H U_0 H^{-1}
U_{\overline{f}}$, where $H$ represents an $n$-bit Hadamard
transform, $U_0$ maps $\ket{0}$ to $-\ket{0}$, and
$U_{\overline{f}}$ maps $\ket{x}$ to $(-1)^{f(x)+1} \ket{x}$. The
algorithm involves applying $G$ repeatedly and observing the
variation in signal intensity. The signal intensity is modulated
in $r$, the number of applications of G, with frequency
proportional to $k$, the number of matches. A more detailed
explanation of this algorithm is found in reference
\cite{jones:counting:1999}.

\begin{figure}
\begin{center}
\subfigure[]{\psfig{file=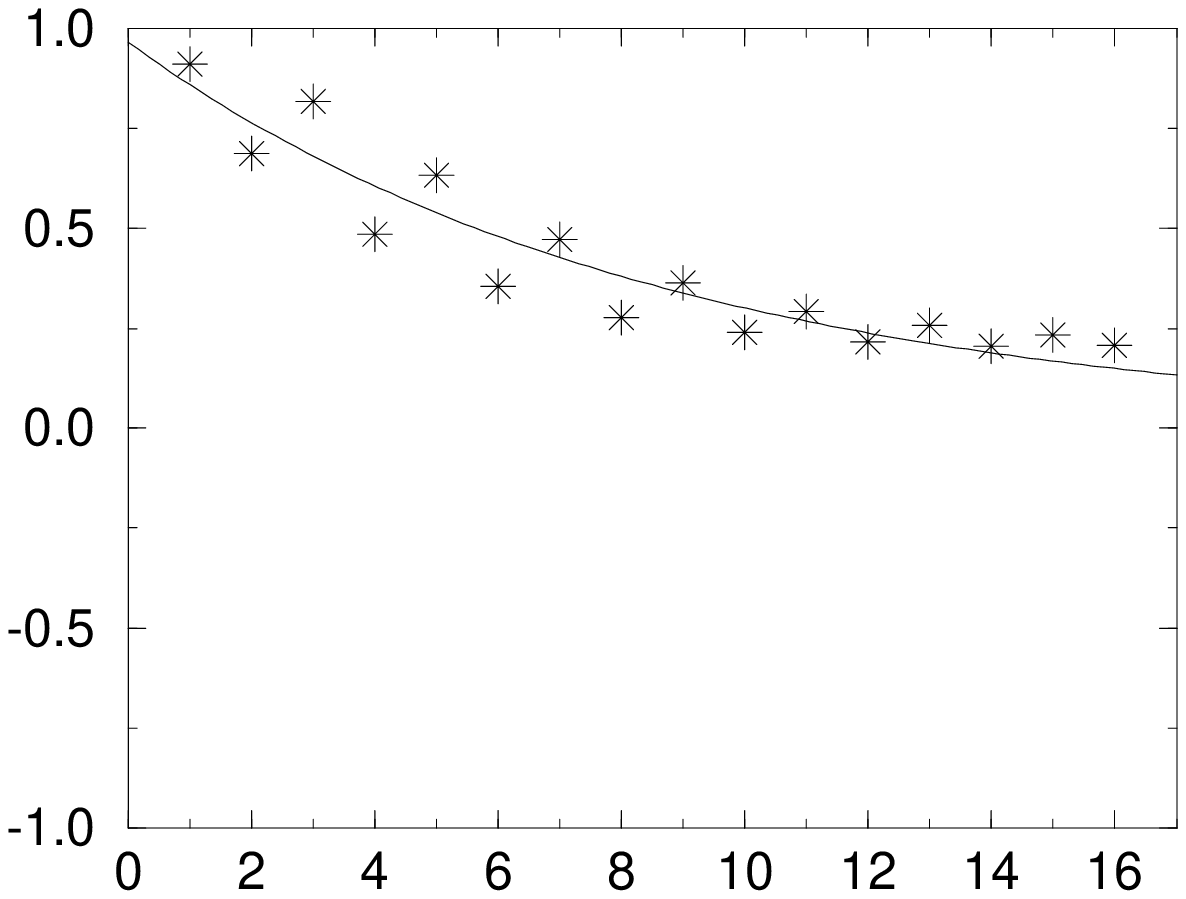,angle=-90,width=\figwidth}}
\qquad
\subfigure[]{\psfig{file=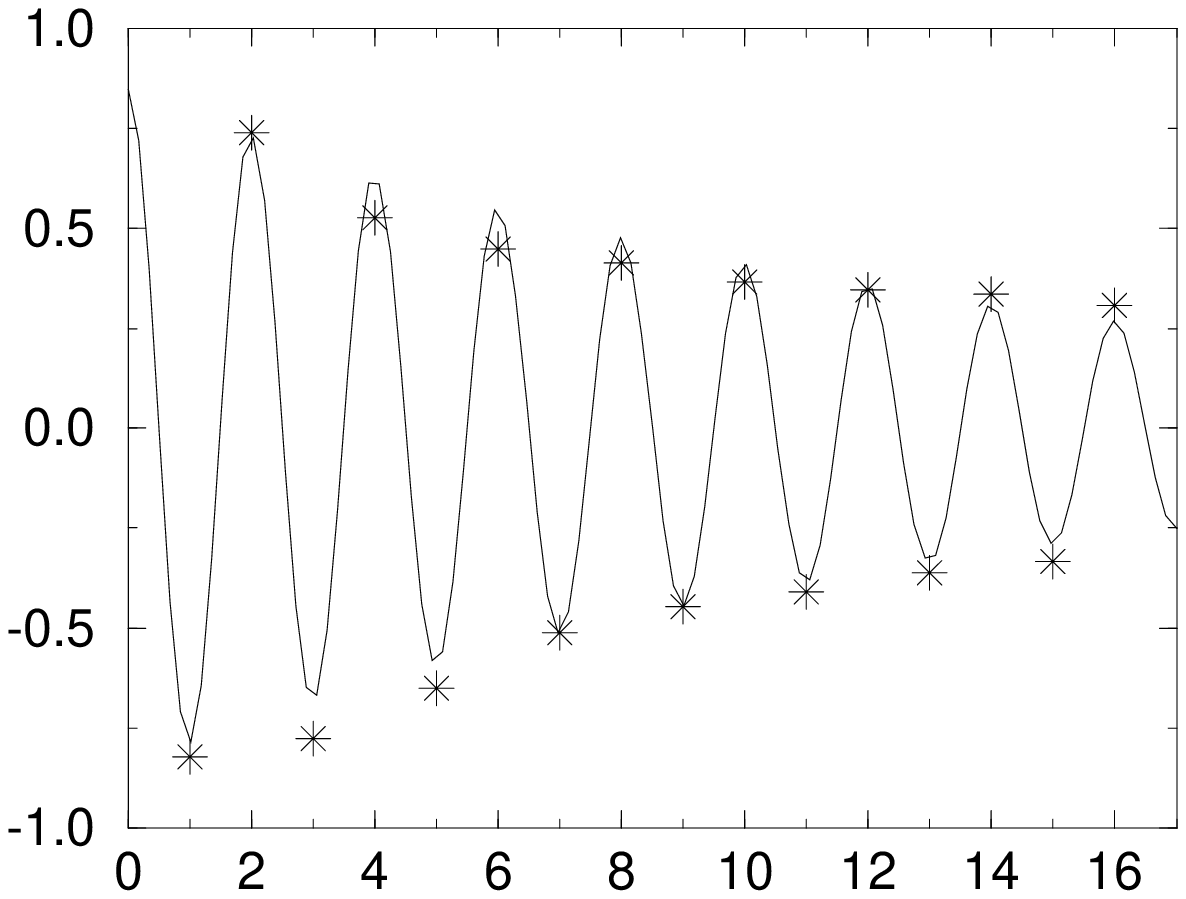,angle=-90,width=\figwidth}}
\end{center} \caption{Experimental results from a 500 MHz NMR
quantum computer for two of the four possible functions $f$: (a)
$f_{00}$ (no matches); (b) $f_{11}$ (both match). The observed
signal intensity is plotted as a function of $r$, the number of
times the controlled-$G$ operator is applied. Intensities are
normalised relative to the case of $r=0$. The solid lines are
exponentially damped cosinusoids with the appropriate theoretical
frequencies, and are plotted to guide the eye.} \label{olddata}
\end{figure}

Figure \ref{olddata} shows experimental results for a search over
a one qubit search space, using a two-qubit quantum computer based
on the two \nuc{1}{H} nuclei of cytosine in solution in
$\mathrm{D_{2}O}$ \cite{jones:1998c}. The NMR spectrometer used a
\nuc{1}{H} operating frequency of 500 MHz. On a one qubit search
space, there are three possible results corresponding to four
possible functions $f$: no inputs match (in which case
$f=f_{00}$),  the first input matches ($f=f_{01}$), the second
input matches ($f=f_{10}$), or both inputs match ($f=f_{11}$). For
simplicity, only the cases $f_{00}$ and $f_{11}$ are shown.

\begin{figure}
\begin{center}
\mbox{\subfigure[]{\psfig{figure=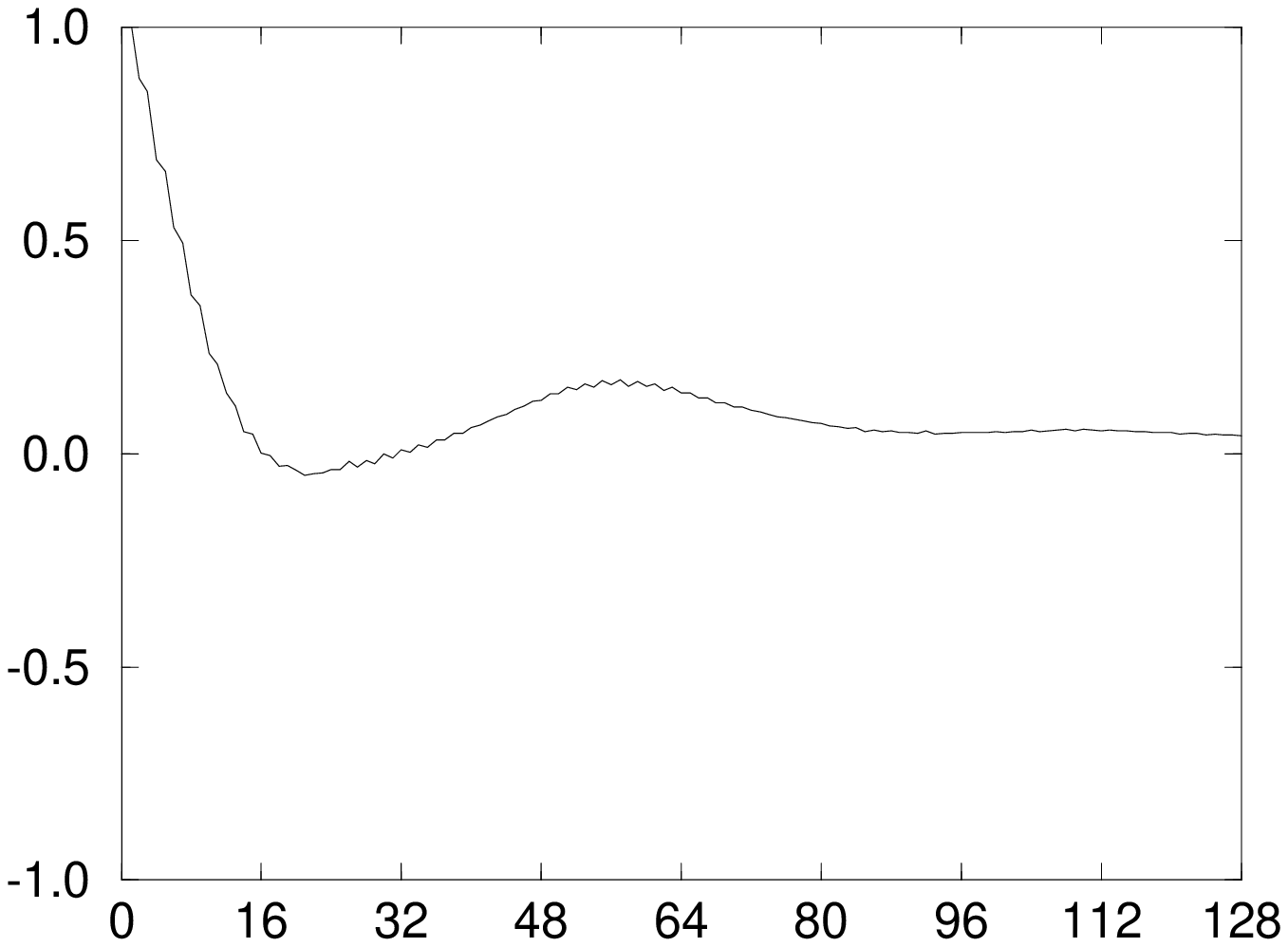,angle=-90,width=\figwidth}}\quad
\subfigure[]{\psfig{figure=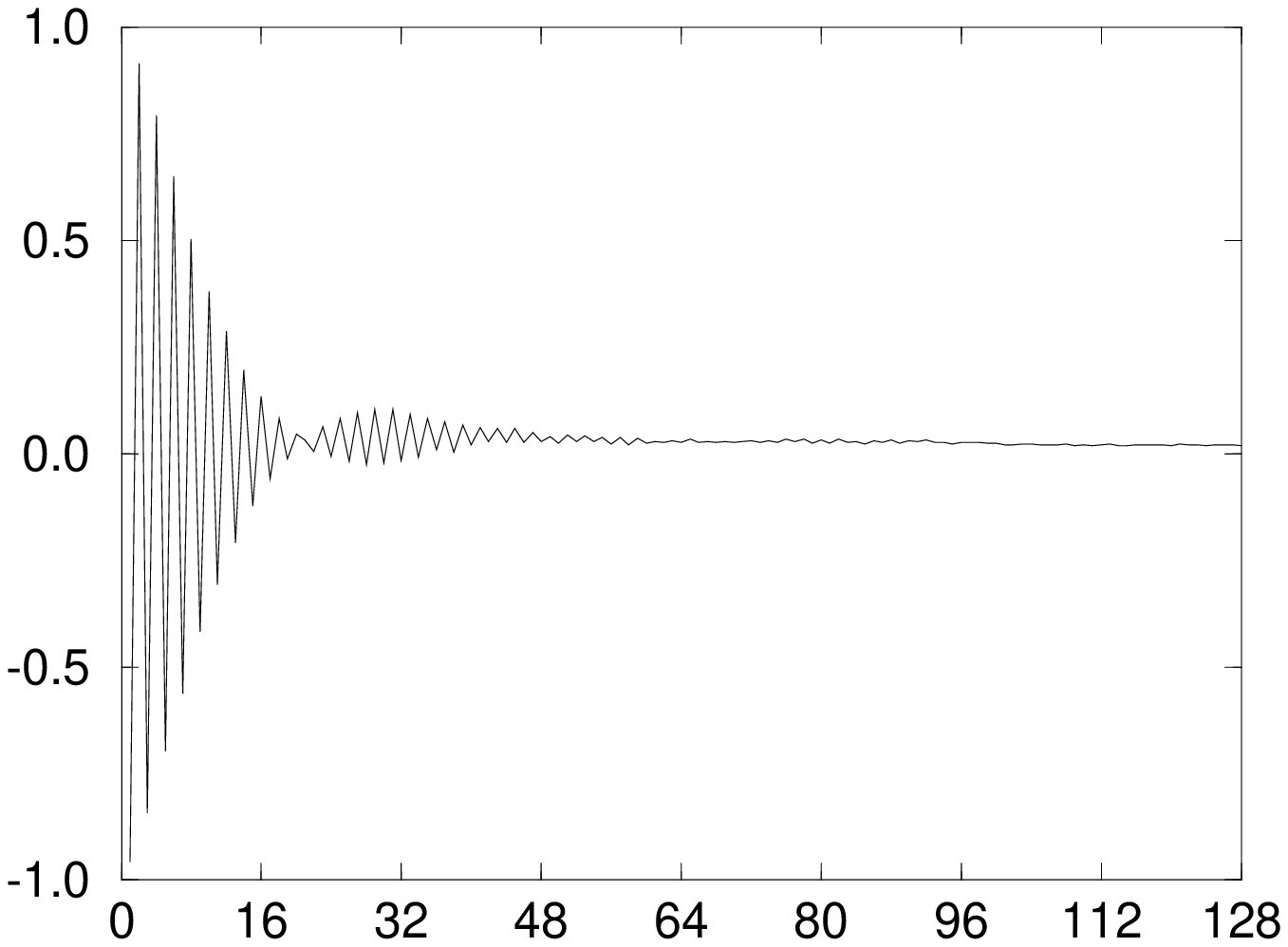,angle=-90,width=\figwidth}}\quad
\subfigure[]{\psfig{figure=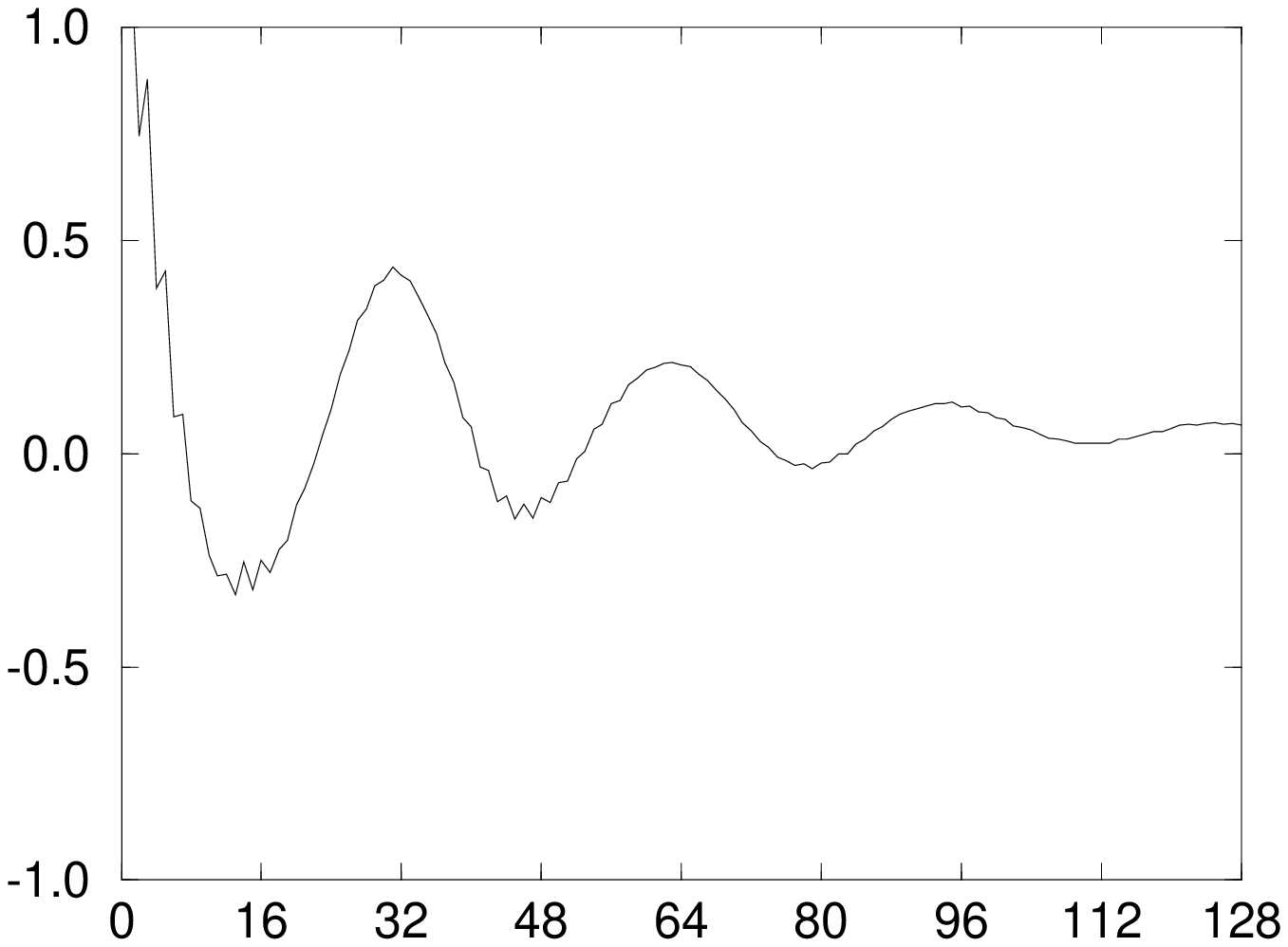,angle=-90,width=\figwidth}}\quad
\subfigure[]{\psfig{figure=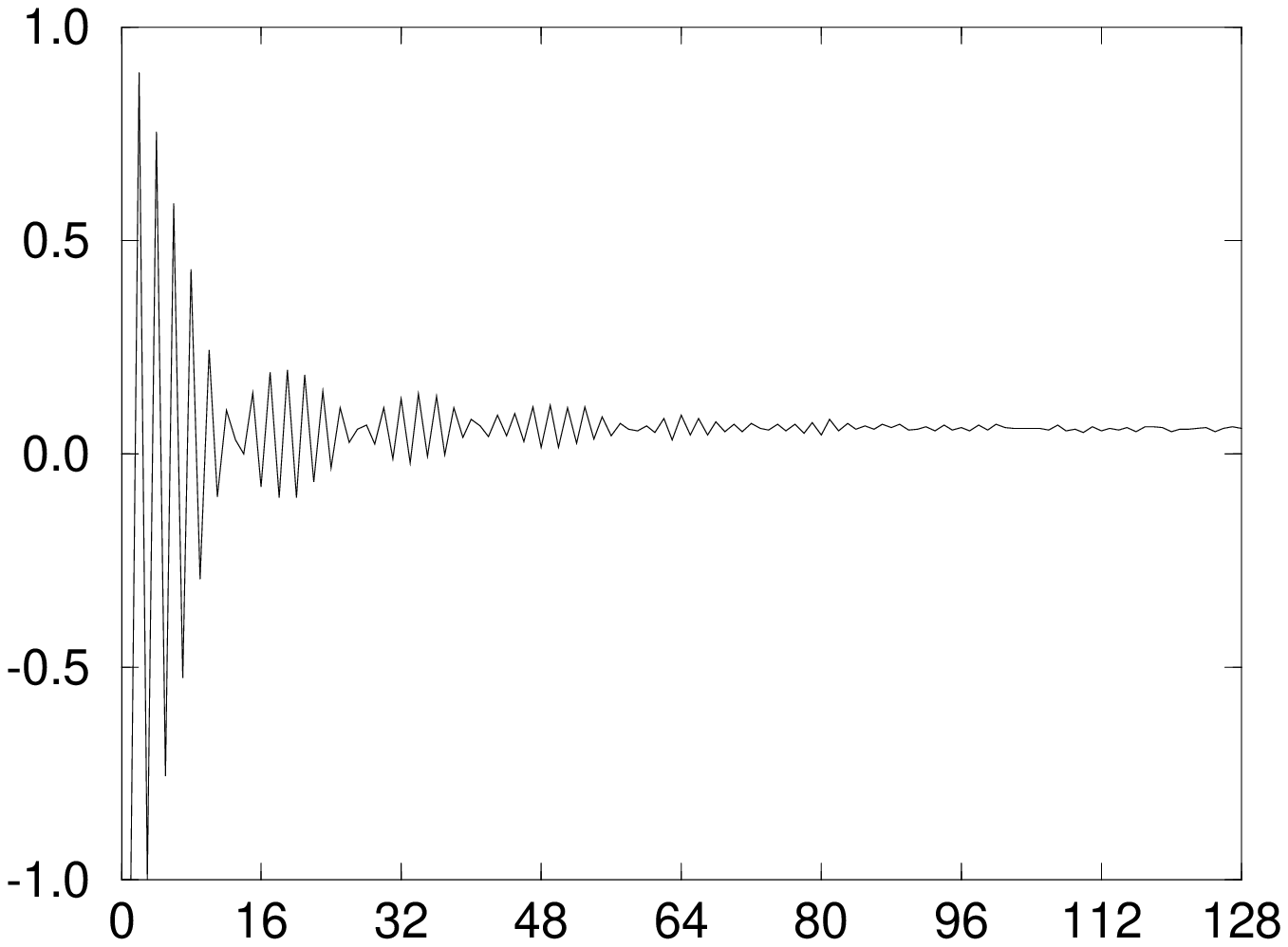,angle=-90,width=\figwidth}}}\end{center}
 \caption{Experimental
results from a 600 MHz ((a) and (b)) and 750 MHz ((c) and (d)) NMR
quantum computer for $f_{00}$ ((a) and (c)) and $f_{11}$ ((b) and
(d) ). The observed signal intensity is plotted as a function of
$r$, the number of times the controlled-$G$ operator is applied.
Intensities are normalized relative to the case of $r=0$. }
\label{600750data}
\end{figure}

The observed signal loss with $r$ could arise from a number of
sources.  Clearly one possibility is the effects of decoherence,
but the observed decay rate is too rapid to be so simply
explained, and patterns in the signal loss (this is especially
clear for $f_{00}$, where data points lie alternately above and
below a smooth curve) suggest a more complex explanation.
Repeating the experiments on spectrometers with higher operating
frequencies markedly degrades the results, as shown in figure
\ref{600750data}, which shows data from computations implemented
on 600 and 750 MHz spectrometers.  Once more, only the $f_{00}$
and $f_{11}$ cases are shown.  These results clearly demonstrate
that the use of expensive high field magnets does not always give
better results.  An unwanted beat frequency and high frequency
chatter have been introduced, and the signal decays more rapidly.
These effects are more severe at 750 MHz than at 600 MHz.

\begin{figure}
\begin{center}
\subfigure[]{\psfig{file=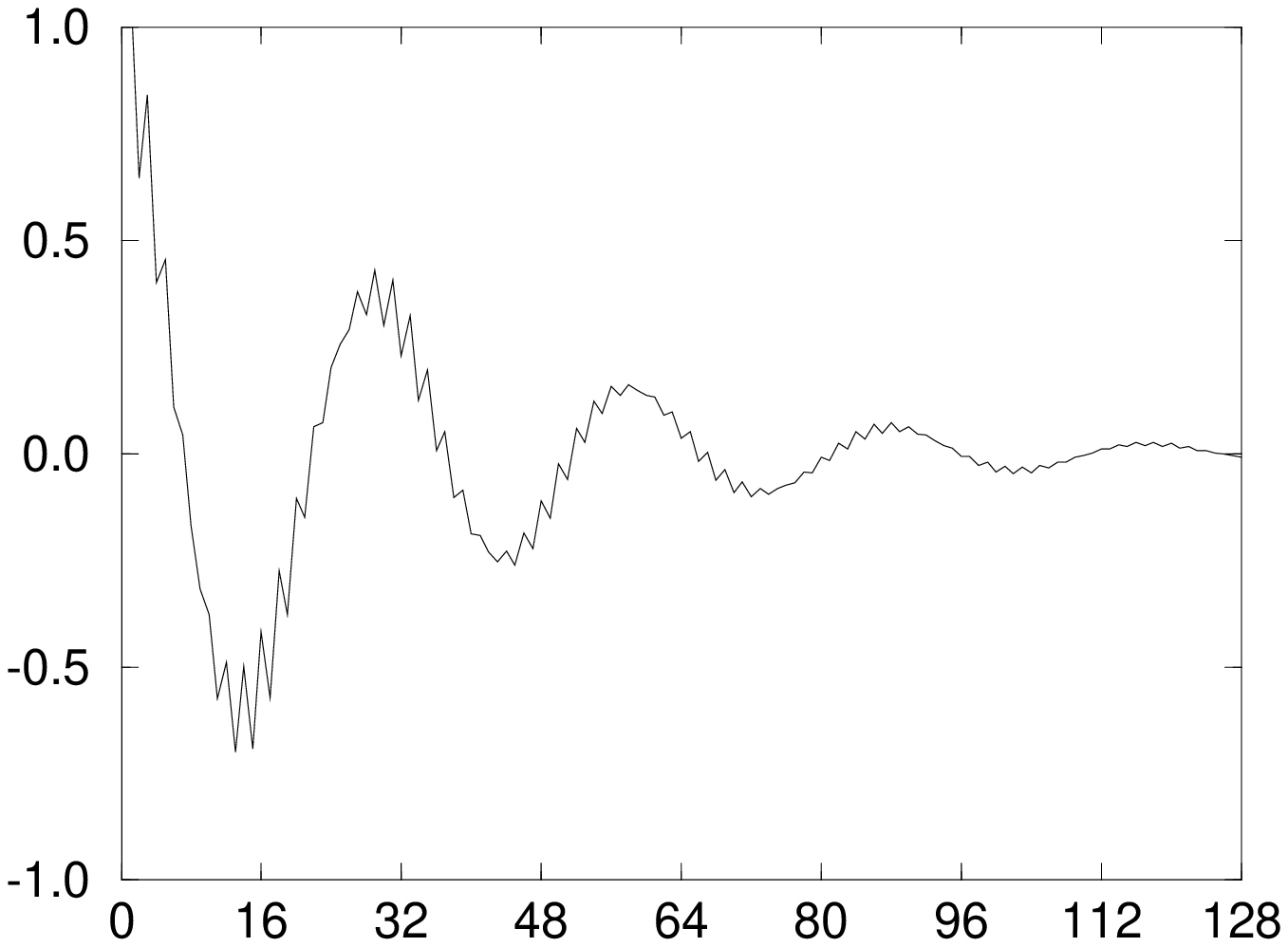,angle=-90,width=\figwidth}}
\qquad
\subfigure[]{\psfig{file=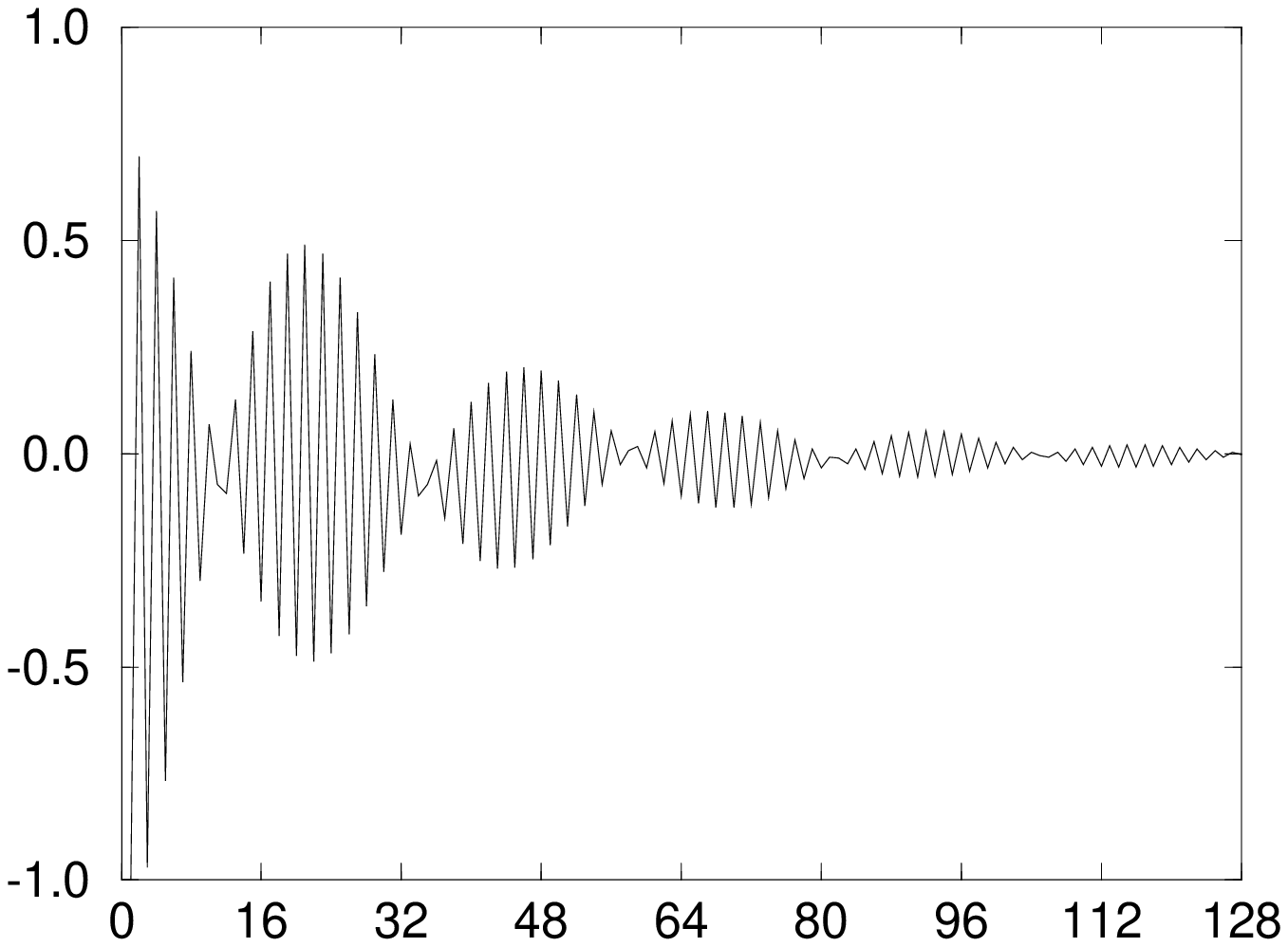,angle=-90,width=\figwidth}}
\end{center}
\caption{Theoretical prediction for 750 MHz spectra once
off-resonance effects are included.  An (arbitrary) exponential
damping has been applied.} \label{bettersim}
\end{figure}

The observed field dependence strongly suggests that these errors
arise from off-resonance effects.  These arise because only a
single RF transmitter is used to excite both \nuc{1}{H} nuclei,
and it cannot be on-resonance with both sets of transitions.
Instead the transmitter is placed halfway between the two
resonance frequencies, resulting in equal and opposite off
resonance terms.  Two factors are responsible for the
poor results at higher frequencies. Firstly, the frequency offsets,
$\pm\delta\nu$, are proportional to the resonance frequency, and
thus are larger at high frequencies. Secondly, the RF field strength
is typically weaker at higher frequencies. These combine to make
the off-resonance effects much more serious. This diagnosis can be
confirmed by including off-resonance effects in simulations of the
NMR experiment. After this correction is made, the theoretical
results, shown in figure \ref{bettersim}, agree strikingly well
with the experimental results. Even the high frequency 'noise' on
the peaks is reproduced.

\begin{figure}
\begin{center}
\subfigure[]{\psfig{file=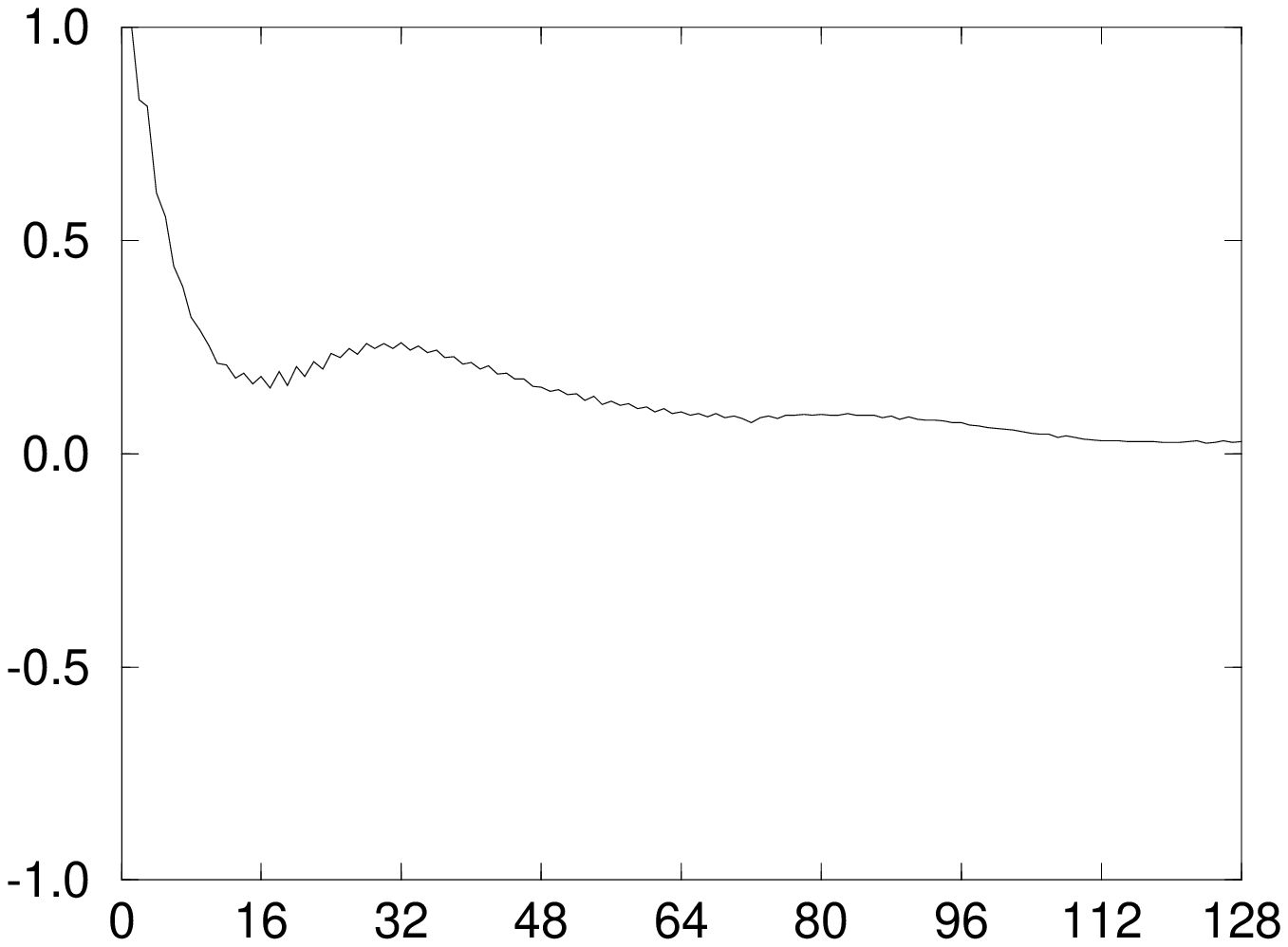,angle=-90,width=\figwidth}}
\qquad
\subfigure[]{\psfig{file=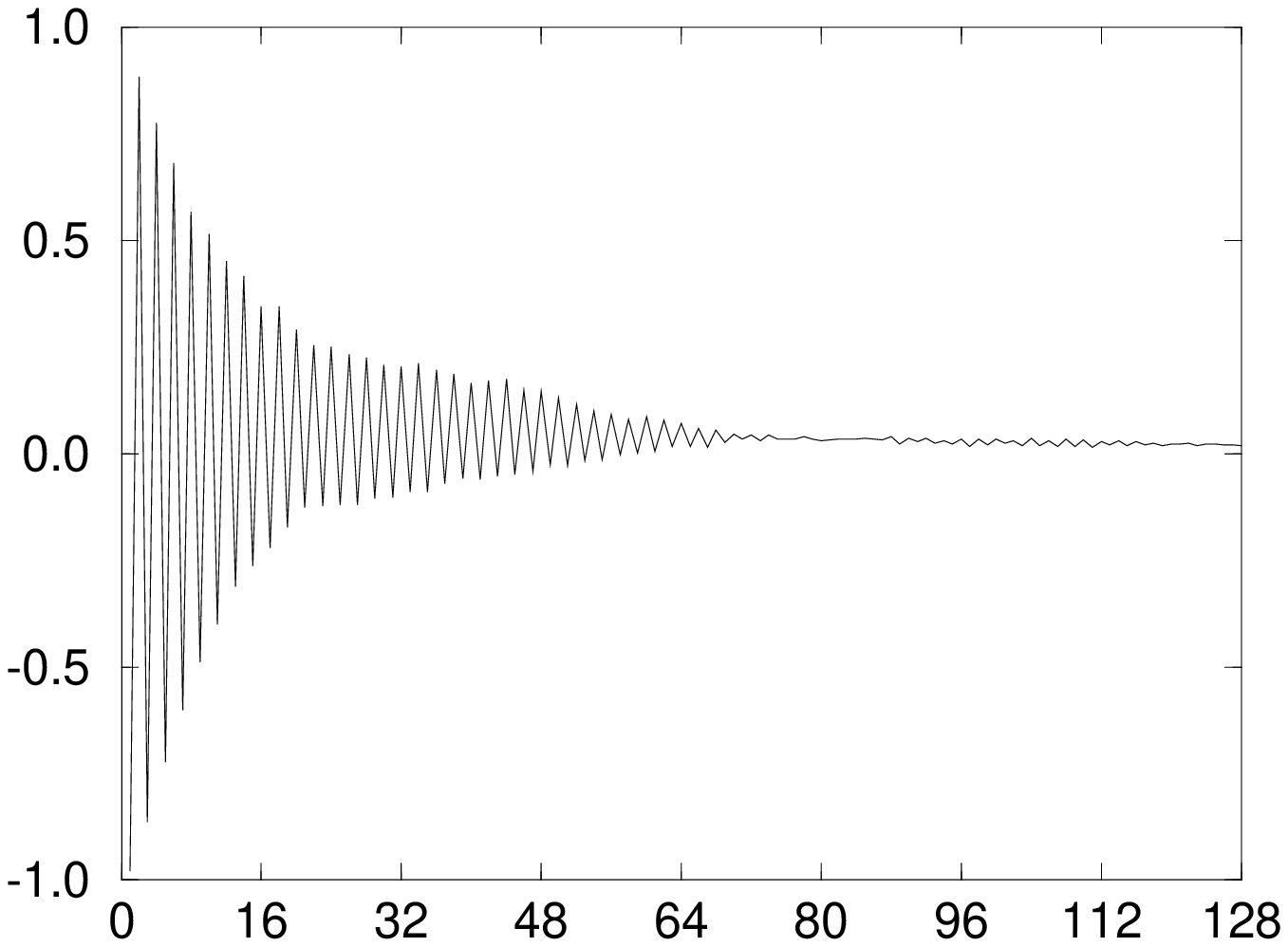,angle=-90,width=\figwidth}}
\end{center}
\caption{Experimental results from the 750 MHz quantum computer
when $90\degree_y$ pulses are replaced by $385_y 320_{-y} 25_y$
pulse sequences.} \label{betterexpt}
\end{figure}

Since our simulations confirm that the major errors in NMR quantum
counting arise from off-resonance effects in the $90\degree_y$
pulses, we chose to use Tycko's $385\degree_y \, 320\degree_{-y}
\, 25\degree_y$ composite $90\degree_y$ sequence, which offers
good compensation for off-resonance errors while remaining fairly
insensitive to pulse-length errors. Substituting this sequence for
the $90\degree_y$ pulses produced a significant improvement in the
experimental results at 750 MHz, as shown in figure
\ref{betterexpt}. The low frequency beating almost disappears, and
the high frequency noise is much reduced.

\begin{figure}
\begin{center}
\subfigure[]{\psfig{file=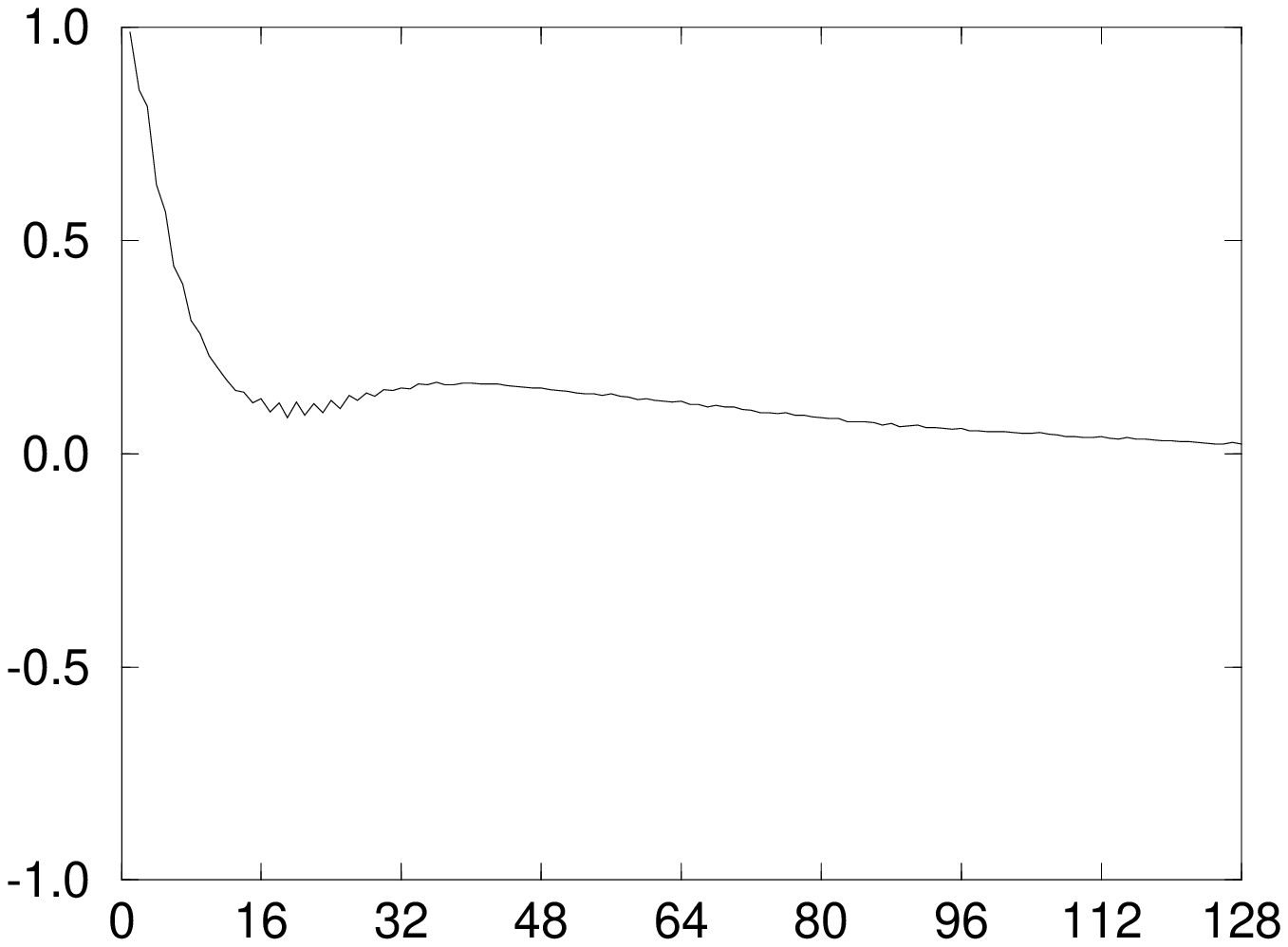,angle=-90,width=\figwidth}}
\qquad
\subfigure[]{\psfig{file=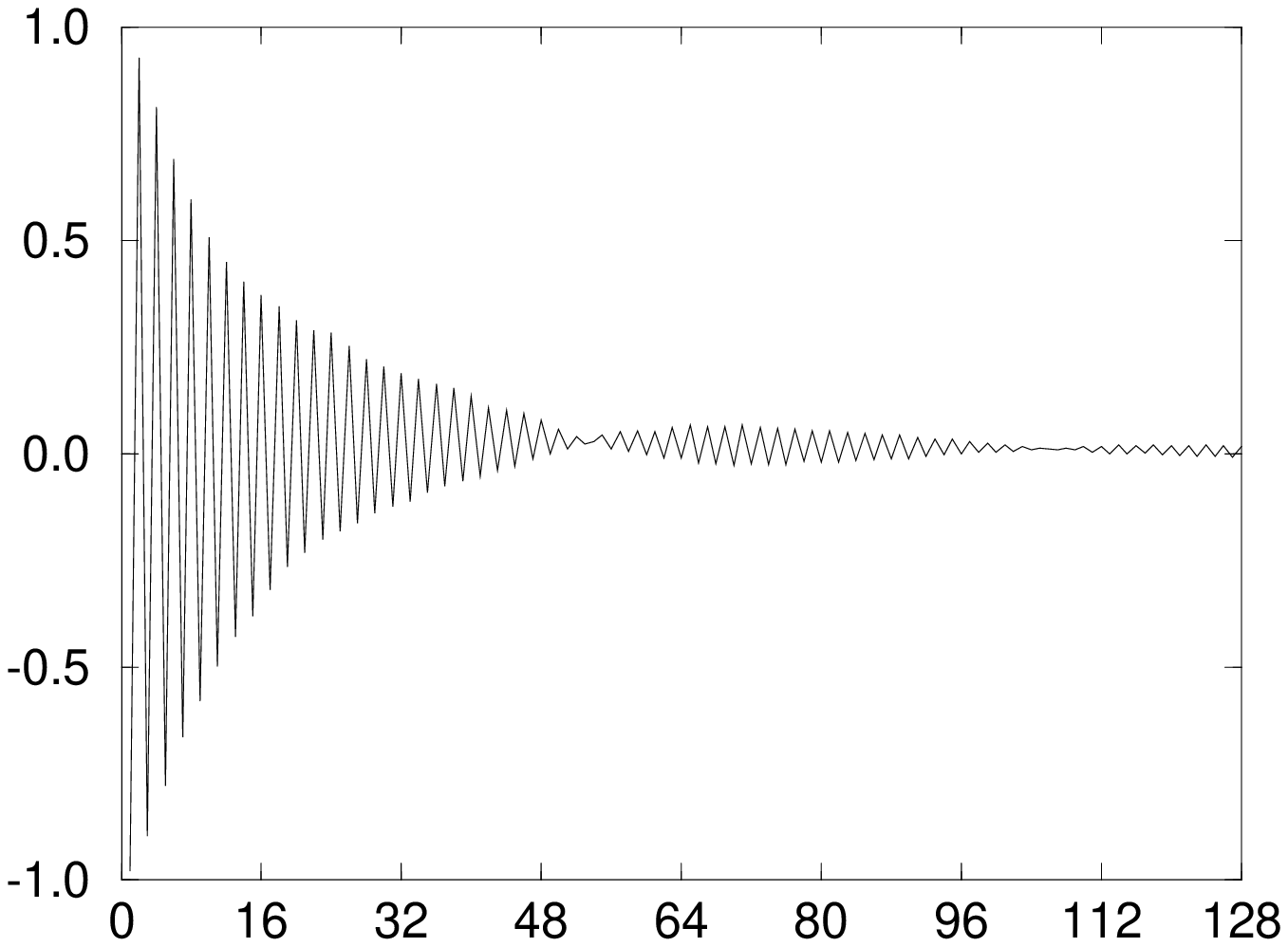,angle=-90,width=\figwidth}}
\end{center}
\caption{Experimental results on the 750 MHz spectrometer when
$90\degree_y$ pulses are replaced by $385\degree_y \,
320\degree_{-y} \, 25\degree_y$ pulse sequences and $180\degree_x$
pulses are replaced by $90\degree_x \, 225\degree_{-x} \,
315\degree_x$ pulse sequences.} \label{with180}
\end{figure}

When using composite pulses, there is a trade-off between the
cancellation of errors and the extra manipulation of the system
necessary to induce this cancellation. While there exist sequences
which in theory compensate for an extraordinary range of errors,
these sequences involve prohibitively long cascades of pulses. In
practice, three-pulse sequences seem to provide a good balance
between insensitivity and simplicity. The hazards of
over-manipulation can be seen when the counting experiment is
repeated using composite $180 \degree $ pulses as well as
$90\degree $ ones, shown in figure \ref{with180}. The composite
$180\degree_x$ pulse used was $90\degree_x \, 225\degree_{-x} \,
315\degree_{x}$ \cite{starcuk:1985}. Although simulations predict
that the results should be slightly better when both $180 \degree$
and $90 \degree$ compensated pulses are used, the results for
$f_{11}$ are in fact slightly worse. The errors introduced by the
extra manipulation outweigh the small gain achieved by the use of
more sophisticated pulse sequences.

\section{Composite pulses for arbitrary rotation angles}
Discussion of composite pulses in the NMR literature has focussed
almost exclusively on $90 \degree$ and $180 \degree$ pulses, for
the simple reason that these are the rotation angles most commonly
used in conventional pulse sequences. While these composite pulses
can be very useful for quantum computation, as shown above, it
would be preferable to have composite pulses for other rotation
angles as well. For example, a $45 \degree$ pulse could be used in
the implementation of a true Hadamard gate \cite{qlgnmr}, which
might be applied many times in a single network.

We use the method of coherent averaging to generate an analytic
expression for composite pulses similar to Tycko's $90\degree$
pulse but with arbitrary rotation angle. The basic idea of the
method is to split the Hamiltonian of the system into intended and
error components,
\begin{equation}
H(t) = H_{0}(t) + V,
\end{equation}
where the ``ideal'' Hamiltonian (written in the rotating frame and
using product operator notation) is
\begin{equation}
H_{0}\parenth{t} = \omega_{1} \parenth{I_x \cos \phi + I_y \sin
\phi}
\end{equation}
and the error term is
\begin{equation}
V=\delta\omega I_z
\end{equation}
($\delta\omega$ is the resonance offset angular frequency).  We
then seek to minimise the propagator of the error component by
expanding it as a power series \cite{tycko:1983}.  Knowing $V$, we
can rewrite the propagator as a product,
\begin{eqnarray}
U(t) & = & U_0(t) U_V(t) \\
 & = & T \exp \parenth{- i \int^\tau _0 dt \, H_0(t)}
 T \exp \parenth{- i \int^\tau _0 dt \, \tilde{V}(t)}
\end{eqnarray}
where $T$ is the Dyson time-ordering operator and
\begin{equation}
\tilde{V}(t) = U_0(t)^{-1} V U_0(t).
\end{equation}
Since $U_V$ varies rather slowly with time, it can be expanded as
a power series in $\tilde{V}$:
\begin{equation}
U_V(\tau)  =  \exp \parenth{-i \tau \parenth{V^{(0)} + V^{(1)} + \dots}}
\end{equation}
using the Magnus expansion
\begin{eqnarray}
V^{(0)} & = & \frac{1}{\tau} \int _0 ^\tau dt \, \, \tilde{V}(t)
\label{vfirstorder}\\
V^{(1)} & = & \frac{-i}{2 \tau} \int _0
^\tau dt \,_1 \int _0 ^{t_1} dt \,_2 \, \brackets{\tilde{V}(t_1),
\tilde{V}(t_2)}\label{vsecondorder}
\end{eqnarray}
etc. Complete expressions for the higher order terms are given in
\cite{haeberlen:1968,haeberlen:1976}. Our objective is to find a
pulse sequence which satisfies the requirement that it perform an
ideal rotation under ideal conditions,
\begin{equation}
T \prod_{i=1}^m U_i(\tau_i) = U_0(\tau)
\label{gendoeswhatitdoes}
\end{equation}
while minimising $V^{(0)}$. We do not consider higher order terms
as using more equations would require more variables in the
solution and we wish to restrict the length of our composite
sequence to three pulses. Here $U_0(\tau)$ is the propagator of an
ideal uncomposed pulse and the $U_i(\tau_i)$ are the constituent
propagators of the ideal composite pulse. Ordinarily, a numerical
search for a solution must be conducted for each set of target
values, $\theta$ and $\phi$, but an analytic solution exists if
the axes of rotation are taken to be $\phi_1 = \phi$, $ \phi_2 =
\pi + \phi$, and $ \phi_3 = \phi$. While this is nominally a loss
of generality, the solutions we find are perfectly compensated to
first order, and so no more general solution would be better to
first order. The solutions also have reasonable properties with
respect to RF field inhomogeneity.

With these phases, it follows trivially from \eq{gendoeswhatitdoes}
that
\begin{equation}
\theta_2 = \theta_1  + \theta_3 - \theta \label{thetaconstraint}
\end{equation}
Working out $V^{(0)}$ with these values of $\phi$ and $\theta_2$
and setting $V^{(0)}_y$ and $V^{(0)}_z$ to zero (it happens that
$V^{(0)}_x$ is already zero) gives a restriction on $\theta_3$,
\begin{equation}
\theta_3 = \pm \theta_1 + 2 n \pi \qquad n = 0, \pm 1, \pm 2, \dots
\label{theta3result}
\end{equation}
Taking the positive result, $\theta_3 = \theta_1 + 2 n \pi $, and
back-substituting gives two solution families,
{\begin{eqnarray}
\theta_1 & = & \pm \arccos \parenth{{\frac{{{\parenth{ 1 - \cos
\theta } }^2} \pm   \sin \theta{\sqrt{\parenth{ 1 - \cos \theta }
\parenth{ 7 +\cos \theta } }} }{4 \parenth{ 1 - \cos \theta } }}}
+ 2 n \pi
\label{vnoughtysoln} \\ \theta_1  & = &  \pm \arcsin
\parenth{\frac{- \sin \theta \parenth{ 1 +  \cos
\theta - \sqrt{\parenth{ 1 + \cos \theta }
\parenth{ 7 +\cos \theta }}}}{4 \parenth{ 1 + \cos \theta } }} + 2 n \pi
\label{vnoughtzsoln}
\end{eqnarray}
} These solutions are equal in the region $ 0 \geq
\theta_{\text{\em target}} \geq \pi$ for the positive root and a
consistent choice of signs. (This is easily checked by working out
the Taylor expansions about 0, and less easily checked by working
out the difference between equations \eqn{vnoughtysoln} and
\eqn{vnoughtzsoln} using an identity for $\arcsin\alpha - \arccos
\beta$.)

Although we have not explicitly considered the effects of RF field
inhomogeneity when determining the $\phi_i$ and $\theta_1$, it
happens they are trivial for the rotation axes we have chosen.
Using $V_{\text{\emph{rf}}} = \delta \omega_{1} \parenth{I_x \cos
\phi
\parenth{t} + I_y \sin \phi \parenth{t}} $, where $\delta
\omega_1$ is a measure of the field inhomogeneity, for $V$ and
working out $V^{(0)}$ with the $\phi_i$ specified above, we get
\begin{eqnarray}
V^{(0)}_x &=& \delta \omega_1\\
V^{(0)}_y &=& 0\\
 V^{(0)}_z &=& 0
\end{eqnarray}
Therefore, equations \eqn{vnoughtysoln} and \eqn{vnoughtzsoln} can
be used exclusively to determine $\theta_1$. We follow Tycko and
choose $n = 1$ so that $2 \pi \leq \theta_1 \leq 4 \pi$. The final
rotation angles are
\begin{eqnarray}
\theta_1 & = & \arccos \parenth{{\frac{{{\parenth{ 1 - \cos \theta
} }^2} +  {\sqrt{\parenth{ 1 - \cos \theta }
\parenth{ 7 +
\cos \theta} }}   \sin \theta}{4 \parenth{ 1 - \cos \theta } }}} +
2 \pi \nonumber\\ \theta_2 & = & \theta_1 + \theta_3 -
\theta \nonumber\\ \theta_3 & = & \theta_1 - 2\pi \label{magictycko}
\end{eqnarray}
Within this family, the quality of the pulses is remarkably
insensitive to to actual details of the pulse sequence. As long as
\eq{thetaconstraint} is strictly satisfied, any sequence with
angles vaguely like those given by \eq{magictycko} will produce
broad-band rotations.

\begin{figure}
\setlength{\unitlength}{\figwidth}
\newsavebox{\axes}
\savebox{\axes}{\put(0,0.09){\makebox(0.12,1)[r]{$\frac{\omega_1}{\omega_1^0}$}}
\put(0.12,0){\makebox(1,0.09)[b]{${\Omega}/{\omega_1^0}$}}}
\centering \mbox{\subfigure[]{
  \begin{picture}(1.1,1.1)
  \put(0,0){\usebox{\axes}}
  \put(0.1,0.1){\psfig{figure=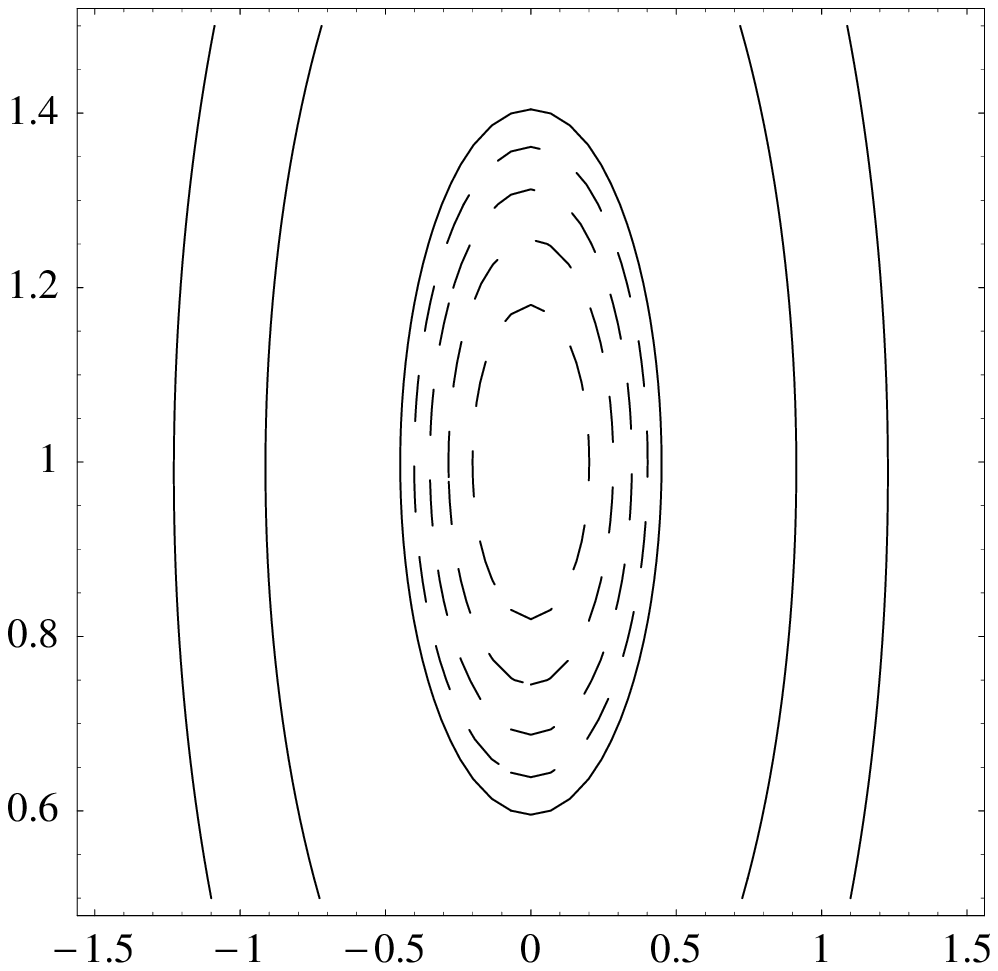,width=\figwidth}}
  \end{picture}
 } \subfigure[]{
  \begin{picture}(1.1,1.1)
  \put(0,0){\usebox{\axes}}
  \put(0.1,0.1){\psfig{figure=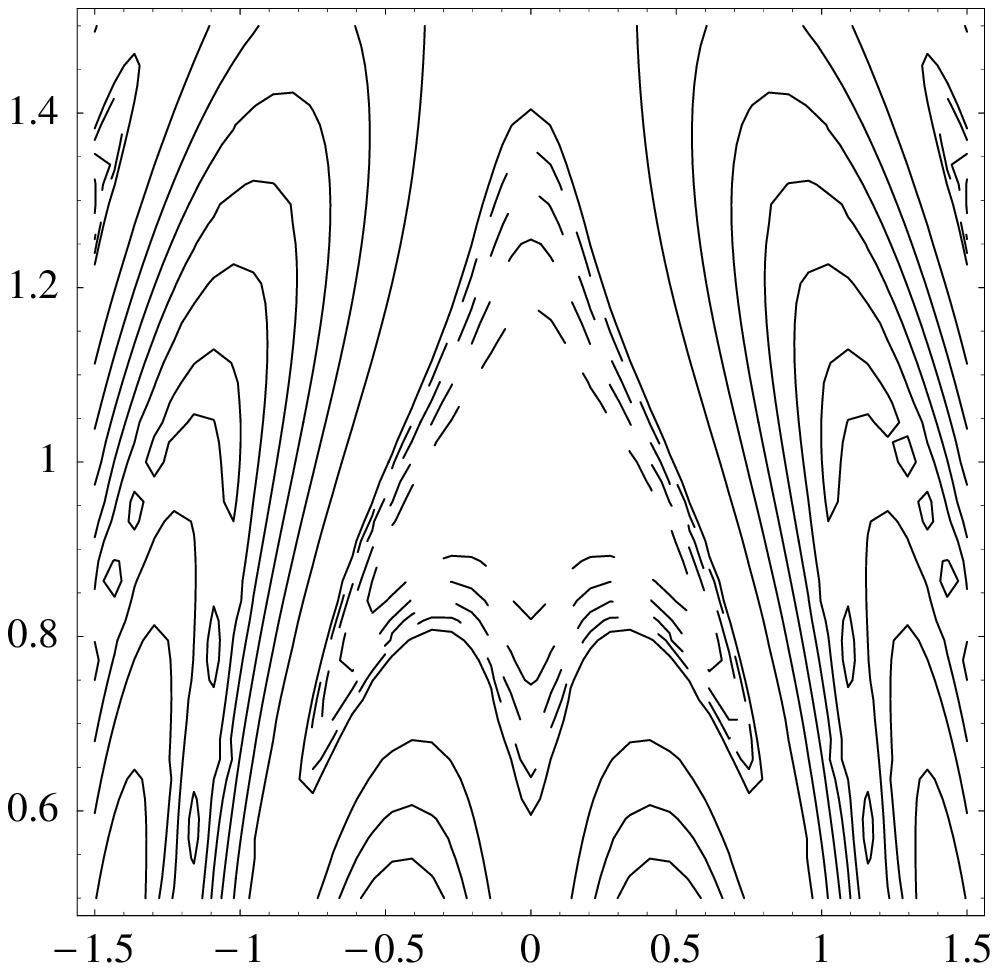,width=\figwidth}}
  \end{picture}
 }}
\mbox{\subfigure[]{
  \begin{picture}(1.1,1.1)
  \put(0,0){\usebox{\axes}}
  \put(0.1,0.1){\psfig{figure=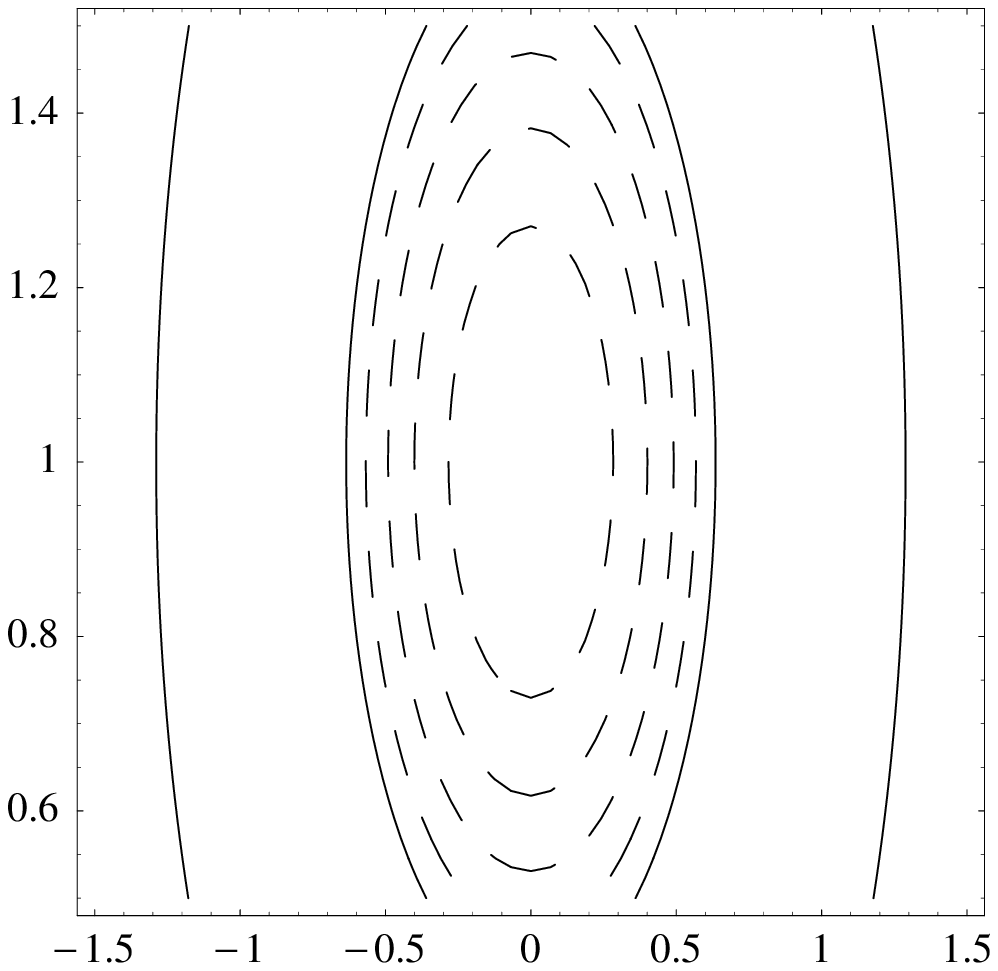,width=\figwidth}}
  \end{picture}
 }\subfigure[]{
  \begin{picture}(1.1,1.1)
  \put(0,0){\usebox{\axes}}
  \put(0.1,0.1){\psfig{figure=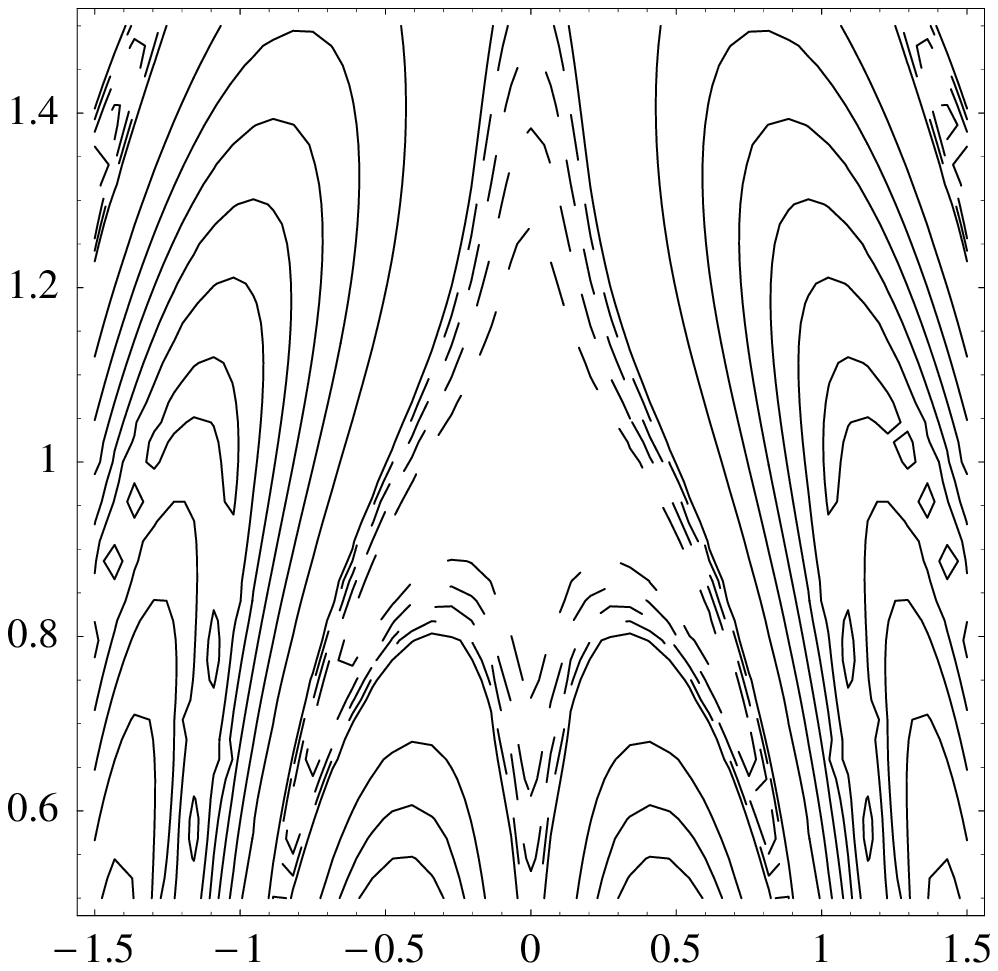,width=\figwidth}}
  \end{picture}
 }}
 \caption{Numerical evaluation of some composite pulses. The
$x$-axis represents radio-frequency inhomogeneity and the $y$-axis
off-resonance effects. The sequences are (a) An uncompensated
$90\degree_x$ pulse (b) Tycko's $90\degree_x$ pulse, $385\degree_x
\, 320\degree_{-x} \, 25\degree_x$ (c) an uncompensated
$60\degree_x$ pulse (d) a new $60\degree_x$ composite pulse,
$375\degree_x \, 331\degree_{-x}  \, 15\degree_{x}$ In all of
these sequence the axes of rotation are $\phi_1 = 0$, $\phi_2 =
\pi$, $\phi_3 = 0$. Solid contour lines are spaced at 15\%
intervals, and dashed contour lines are spaced at 1\% intervals,
beginning at 95\%.} \label{merits}
\end{figure}

Figure \ref{merits} shows fidelity plots for Tycko's original $90
\degree$ composite pulse, and a new pulse derived using our
analytic method. For comparison purposes, fidelity plots for the
equivalent uncompensated pulses are included as well. The fidelity
was evaluated using quaternions as described in reference
\cite{levitt:1986}.

\begin{figure}
\centering
\mbox{\subfigure[]{\psfig{figure=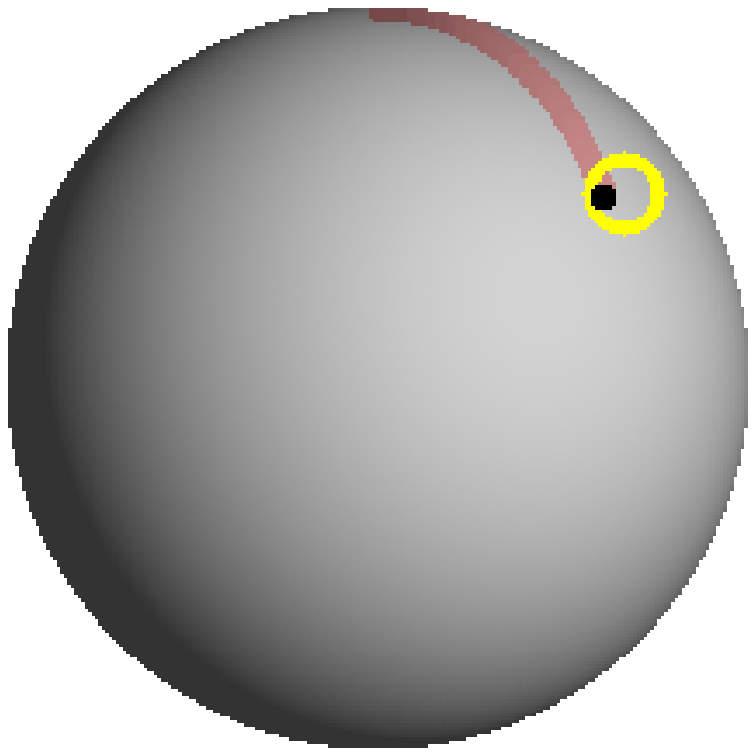,width=\figwidth}}\qquad
\subfigure[]{\psfig{figure=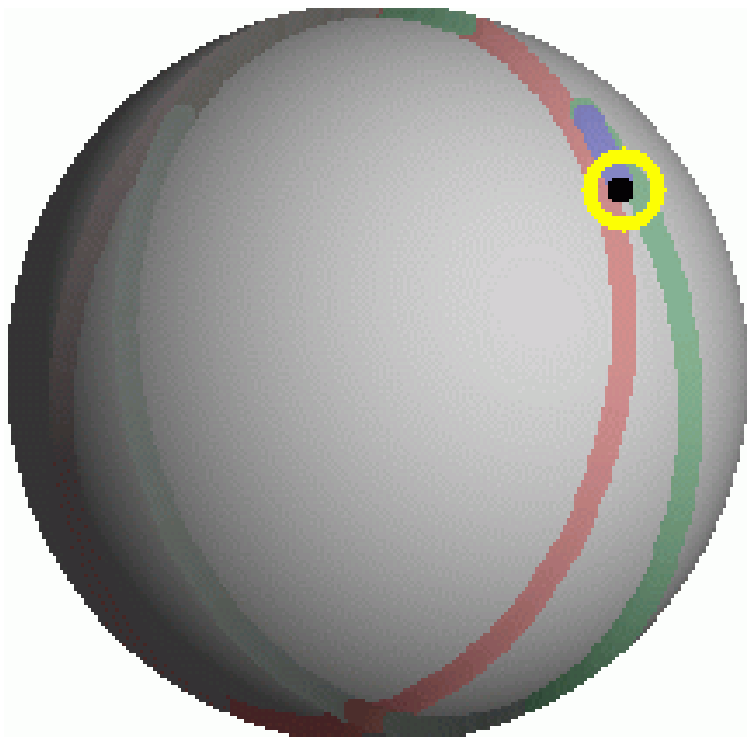,width=\figwidth}}}
\caption{Magnetization trajectories on the Bloch sphere for a
$60\degree$ uncompensated pulse and Tycko-type composite pulse for
a system with a fairly large resonance offset. The black dot shows
the final position on the Bloch sphere, and the pale ring shows
the target position. The three rotations within the Tycko-type
sequences are shown in different colours. (a) An uncompensated
rotation of a spin system with a resonance offset of 2 ppm in a
750 MHz magnet with an RF field strength of 10 KHz. (b) A
composite rotation in the same system.} \label{grapefruit}
\end{figure}

The operation of the sequence whose fidelity is plotted in Figure
\ref{merits}(d) is illustrated in Figure \ref{grapefruit}. The
resonance offset was chosen to be fairly large so that the
mechanism of the composite pulse is clear.

The composite pulses we have derived have good properties in both
quantum computation and spectroscopic contexts, but it may be
possible to produce pulses with properties that are even more
favourable for quantum computation if we stop using spectroscopy
as our implicit model. Consider \eq{gendoeswhatitdoes} again.
Although it seems like an obvious condition that the behaviour of
the pulse should be ideal in the absence of errors, this is
misleading. We would never want to use a composite pulse in a
situation where all pulses were on-resonance, and so there is no
\emph{a priori} reason to design pulses which perform well in the
event of this non-existent ideality. The resonance offsets for
quantum computation will always be known exactly, so we would like
to be able to design pulse sequences which perform best as these
known offsets rather than at the zero offset. This is a problem we
are currently considering.

\section{Conclusions}
Composite pulses show great promise for reducing data errors in
NMR quantum computers. More generally, any implementation of a
quantum computer must be concerned on some level with rotations on
the Bloch sphere, and so composite pulse techniques may find very
broad application in quantum computing. Composite pulses are not,
however, a panacea, and some caution must be exercised in their
use.

\acknowledgments JAJ is a Royal Society University Research
Fellow. HKC thanks NSERC (Canada) and the TMR programme (EU) for
their financial assistance.  We thank Starlab (Riverland NV) for
their support. The OCMS is supported by the UK EPSRC, BBSRC, and
MRC.

\end{document}